\documentclass[11pt]{article}
\usepackage{rotating,amsmath,amssymb,amsfonts,amsthm,longtable,caption,url}
\usepackage{epsfig,mathrsfs,natbib,bm,color,graphics,multirow,makecell,booktabs}
\usepackage{multirow}
\usepackage{adjustbox}
\usepackage{graphicx}
\usepackage[ruled,vlined]{algorithm2e}
\def\boxit#1{\vbox{\hrule\hbox{\vrule\kern6pt \vbox{\kern6pt#1\kern5pt}
\kern6pt\vrule}\hrule}}

\newcommand{\indep}{\rotatebox[origin=c]{90}{$\models$}}

\newcommand{\bX}{{\boldsymbol X}}

\newtheorem{theorem}{Theorem}[]

\newtheorem{assumption}{Assumption}[]

\newtheorem{remark}{Remark}[]
\def\ANNALS{{\it Annals of Statistics}}

\def\JRSSB{{\it Journal of the Royal Statistical Society, Series B}}

\def\JASA{{\it Journal of the American Statistical Association}}

\def\JASA{{\it Journal of the American Statistical Association}}
\def\ANNALS{{\it Annals of Statistics}}

\def\JRSSB{{\it Journal of the Royal Statistical Society, Series B}}

\def\ANNALS{{\it Annals of Statistics}}

\def\JRSSB{{\it Journal of the Royal Statistical Society, Series B}}

\def\JASA{{\it Journal of the American Statistical Association}}

\def\PNAS{{\it Proceedings of the National Academy of Sciences USA}}

\date{}
\begin{document}

\title{A Double Regression Method for Graphical Modeling of  High-dimensional Nonlinear and Non-Gaussian Data}

%\author{Siqi Liang and Faming Liang}

 \author{Siqi Liang and Faming Liang\thanks{
    To whom correspondence should be addressed: Faming Liang.
    The authors gratefully acknowledge the support in part to this work by the NSF grant DMS-2015498 and the NIH grant R01GM126089, and thank the editor, associate editor and referees for their constructive comments which have led to significant improvement of this paper.}\hspace{.2cm}\\
    Department of Statistics \\
    Purdue University}

\maketitle

\begin{abstract}
 Graphical models have long been studied in statistics as a tool for inferring conditional independence relationships among a large set of random variables. The most existing works in graphical modeling focus on the cases that the data are Gaussian or mixed and the variables are linearly dependent. In this paper, we propose a double regression method for learning graphical models under the high-dimensional nonlinear and non-Gaussian setting, and prove that the proposed method is consistent under mild conditions. The proposed method works by performing a series of nonparametric conditional independence tests. The conditioning set of each test is reduced via a double regression procedure where a model-free sure independence screening procedure or a sparse deep neural network can be employed. The numerical results indicate that the proposed method works well for high-dimensional nonlinear and non-Gaussian data. 

\vspace{2mm}


\end{abstract}

\section{Introduction}

Graphical models, which generally refer to undirected Markov networks,  have proven to be a useful tool for inferring conditional independence relationships for a large set of random variables. They are particularly useful under the high-dimensional scenario, i.e., when the number of random variables is greater than the number of observations. 
In this scenario, with the aid of the sparse graphical model learned for the explanatory variables, the inference for high-dimensional regression is reduced to the inference for a series of low-dimensional regression, for which the $t$-test can be used for inference of the significance and associated confidence interval for each explanatory variable.
See \cite{LiangXJ2021} for the details.  
%In this scenario, the graphical models also serve the purpose of dimension reduction in statistical modeling of the data so that probabilistic inference queries can be made efficiently. 
%As summarized in \cite{Tillman2009}, the graphical models generally serve two purposes: (i) finding compact representations of probability distributions so that probabilistic inference queries can be made efficiently, and (ii) modeling unknown data generating mechanisms and predicting causal relationships.
As an alternative to undirected Markov networks, 
Bayesian networks can also serve the purpose of inference for the conditional independence relationships for a large set of random variables. However, Bayesian networks are directed and focus more on data generating mechanisms. It is worth noting that the Bayesian network can be constructed from its moral graph, which is an undirected Markov network, using a collider set algorithm \citep{PelletE2008} or local neighborhood algorithm \citep{MargaritisT2000} to have the edge directions added. Therefore, learning undirected Markov networks is fundamental, which is the focus of this paper.

%As summarized in \cite{Tillman2009}, they generally serve two purposes: (i) finding compact representations of probability distributions so that probabilistic inference queries can be made efficiently, and (ii) modeling unknown data generating mechanisms and predicting causal relationships.

In practice, the random variables can be Gaussian or non-Gaussian, and the dependence between different random variables can be linear or nonlinear. However, 
most of the existing works in graphical modeling focus on Gaussian data with linear dependence.
Under the Gaussian and linear assumptions, various methods have been developed 
 based on the properties of partial correlation coefficients, concentration matrix, and regression coefficients of Gaussian graphical models (GGMs), see e.g., graphical Lasso \citep{YuanL2007, FriedmanHT2008}, nodewise regression \citep{MeinshausenB2006},  $\psi$-learning \citep{LiangSQ2015}, and structural  equation modeling \citep{GGMnotears2018}.
 To be more detailed, let  $\{X_1, X_2, \ldots, X_{p}\}$ denote a set of zero-mean Gaussian random variables, let $V=\{1,2,\ldots,p\}$ denote its index set, let $\Sigma$ denote the covariance matrix of the random variables, let $C=\Sigma^{-1}$ denote the concentration matrix, and let $\beta_i^{(j)}$ denote a coefficient of the normal linear regression 
 \begin{equation} \label{SEMeq}
 X_j=\beta_i^{(j)} X_i+\sum_{r \in V\setminus \{i,j\}} \beta_r^{(j)} X_r+\epsilon_j, \quad j=1,2,\ldots,p,
 \end{equation} 
 where $\epsilon_j$ is a zero-mean Gaussian random error.  By the standard theory on normal linear regression, see e.g. chapter 13 of \cite{Buhlmann2011Book},  the following relation holds:
 \begin{equation} \label{Gaussianrelationeq}
 \rho_{ij|V\setminus \{i,j\}} \ne 0 \Longleftrightarrow C_{ij}\ne 0
 \Longleftrightarrow  \beta_i^{(j)} \ne 0,
 \end{equation}
 where $\rho_{ij|V\setminus \{i,j\}}$ denotes the 
 partial correlation coefficient of $X_i$ and 
 $X_j$ conditioned on all other variables, and 
 $C_{ij}$ denotes the $(i,j)$th element of the concentration matrix $C$. 
 Based on this relation,  the graphical Lasso method \citep{YuanL2007,FriedmanHT2008} 
 infers the GGM by directly estimating the concentration matrix via a regularization approach; 
 the nodewise regression method \citep{MeinshausenB2006} infers the GGM by computing the coefficient $\beta_i^{(j)}$ for each pair of variables $(X_j, X_i)$ via a regularized regression;  
the graphical regression method \citep{ni2019bayesian, zhang2020gaussian} regresses both the mean and the precision matrix of a GGM on covariates, which enables estimation of subject-specific graphical models; the $\psi$-learning method \citep{LiangSQ2015} infers the GGM by an equivalent measure of the partial correlation coefficient $\rho_{ij|V\setminus \{i,j\}}$, which can be computed based on a reduced conditioning set; and  
the structural equation modeling method \citep{GGMnotears2018} infers a directed acyclic graph (DAG) for the GGM by directly solving the structure equations formed by the nodewise regression (\ref{SEMeq}), where the problem was formulated as a continuous optimization problem with a structural constraint \citep{HararyM1971} for ensuring acyclicity of the graph.   
 
 Many authors have tried to extend the methods developed for GGMs to non-Gaussian data or mixed data under the assumption that the random variables are still linearly dependent. For example, \cite{RavikumarWL2010} extends the nodewise regression method to binary data; \cite{LeeHastie2013}, \cite{Chengetal2013} 
  and \cite{Yangetal2014} extends the nodewise regression method to  mixed data; 
 \cite{XuJiaLiang2019} extends the $\psi$-learning method to mixed data; and \cite{FanLY2017} extends  the graphical Lasso method to mixed data by 
 introducing some latent variables for the observed discrete  variables. However, as pointed out in \cite{FanLY2017}, the latent variable method does not necessarily lead to correct graphical models, as the conditional independence relationship between the latent variables does not imply the conditional independence relationship between the observed discrete variables. The copula PC method \citep{CuiGH2016}  suffers from the same issue. 
 Quite recently,  \cite{GGMnotears_nonlinear2020} extends the structural equation modeling method to certain types of nonlinear models defined in the Sobolev space. However, it is unclear whether the method can lead to consistent estimation of the underlying graphical models under the high-dimensional scenario. Some other methods,  e.g. \cite{Yu2019DAGGNNDS}, infer the model in an approximate sense using variational autoencoder, and it is unclear whether it can lead to the conditional independence relationships pursued in graphical modeling.  
 
 Up to our knowledge, none of the works has been done for inference of graphical models under the general high-dimensional nonlinear and non-Gaussian setting. 
 To tackle this problem, we proposed a double regression method. The proposed method works by performing a series of nonparametric conditional independence tests, for each of which the conditioning set is reduced by a double regression procedure. The proposed method is 
 shown to be consistent under mild conditions. 
 The numerical results indicate that the proposed method works well for high-dimensional nonlinear and non-Gaussian data. 
 
 \textcolor{black}{Statistical learning for nonlinear and non-Gaussian models is important, as in modern data science nonlinear dependency is common and the data may not tend towards Gaussian distributions, such as biological data and financial data. Researchers have found that although the accuracy of the partial correlation test-based GGM methods are not significantly affected by violations of the Gaussian assumption, they are significantly affected by violations of the linear dependency assumption, see e.g. \cite{voortman2008insensitivity} for 
 more discussions on this issue. 
 Researchers have also found that nonlinearity and non-Gaussianity can actually be a blessing, and nonlinear non-Gaussian modeling can reveal more accurate information about the true data generating process than the linear and Gaussian approximation. For example, \cite{Shimizu2006ALN} and \cite{Hoyer2008CausalDO} found that non-Gaussianity is helpful in predicting causal relationships among the variables. }
  
The remaining part of this article is organized as follows. Section \ref{doubleregressionmethod} describes the proposed method and establishes its consistency. Section \ref{experiments} illustrates the proposed method using simulated data. Section \ref{sect4} applies the proposed method to identification of drug-sensitive  genes with the cancer cell line encyclopedia (CCLE) data. Section \ref{computationalcomplexanalysis} analyzes the computational complexity of the proposed method. Section \ref{disucssion} concludes the paper with a brief discussion. 
 
 %\section{Existing Work} 
 
 %\subsection{SEM, notears}  
 %\cite{GGMnotears2018}
 %\subsection{Nonlinear SEM1}
 %\cite{GGMnotears_nonlinear2020}
 %\subsection{Nonlinear/non-Gaussian SEM 2} 
 %\cite{Yu2019DAGGNNDS}

 \section{Double Regression Method} \label{doubleregressionmethod}
 
 \subsection{The Algorithm} 
 
  {\it Notations:} Consider a set of random variables $\{X_1, X_2, \ldots, X_{p}\}$, where each variable can be non-Gaussian and the dependence between different variables can be nonlinear. The dimension $p$ is assumed to grow with the sample size $n$. To indicate this dependence, we rewrite $p$ as $p_n$ in what follows. 
  Let $V=\{1,2,\ldots,p_n\}$ be the index set of the variables,  
  let $A\subset V$ be a subset of $V$, and let $\bX_A=\{X_k: k \in A\}$.  
  
  As our goal is to construct a graphical model for 
  the variables, the conditional independence test (CIT) $X_i \indep X_j|X_{V\setminus \{i,j\}}$ needs to be conducted for each pair of the variables $(X_i, X_j)$. Since the functional form of the dependence between different variables is unknown, a nonparametric CIT can be applied here. An abundance of nonparametric CITs have been developed in the literature. A non-exhaustive list includes permutation-based tests \citep{Doran2014,Berrett2019}, kernel-based tests  \citep{ZhangKCIT2012,Strobl2019}, classification or regression-based tests \citep{Sen2017,Zhang2017FeaturetoFeatureRF},  knockoff-based tests \citep{knockoffs2018}, 
  and generative adversarial network (GAN)-based tests
  \citep{bellot2019CItest}. Refer to \cite{ChunF2019} for a comprehensive overview.

% There are many choices for the nonparametric CIT. For example, if the variables are continuous, the kernel CIT  \citep{Strobl2019,ZhangKCIT2012}
  %available at {\tt https://rdrr.io/github/ericstrobl/RCIT/}, 
  %and model-powered CIT \citep{Sen2017} can be applied. 
 % Refer to \cite{ChunF2019} for an overview.  
 % For more general case, the generative CIT \cite{bellot2019CItest} can be used. 

As pointed out in \cite{ChunF2019}, the existing nonparametric CITs often suffer from the curse of dimensionality in the confounding set; that is, the tests may be ineffective when the sample size is small, since the accumulation of spurious correlations from a large number of confounding variables makes them difficult to discriminate between the hypotheses. To tackle this issue, we consider the following simple mathematical fact of conditional probability distributions:
  \begin{equation} \label{prob1} 
  \begin{split}
  P(X_i, X_j|\bX_{V\setminus \{i,j\}}) & = 
  P(X_i|X_j, \bX_{V\setminus \{i,j\}})
   P(X_j| \bX_{V\setminus\{i,j\}}) 
   =  P(X_i|X_j, \bX_{S_i\setminus\{j\}}) P(X_j|\bX_{S_{j\setminus i}}) \\
 &  = P(X_i, X_j|\bX_{S_i\setminus\{j\}}, \bX_{S_{j\setminus i}})=P(X_i,X_j|\bX_{S_i\cup S_{j \setminus i}\setminus\{j\}}),
 \end{split}
  \end{equation} 
  where $\bX_{S_i}$ denotes the set of true variables of the nonlinear regression $X_i \sim \bX_{V\setminus\{i\}}$, and 
  $\bX_{S_{j\setminus i}}$ denotes the set of true variables of the nonlinear regression $X_j \sim \bX_{V\setminus\{i,j\}}$.
  Based on this property of conditional probability distributions, we propose the double regression method which is summarized in Algorithm \ref{Alg1} and can be used for reducing the 
  conditioning sets of the CITs.    
 
\begin{algorithm}[htbp] 
   \SetAlgoLined
        \For{each variable $X_i$, $i=1,2,\ldots,p_n$}{
 \textbf{(i) Variable selection:} conduct nonlinear/non-Gaussian regression 
        \begin{equation} \label{screeneq1}
         X_i\sim \bX_{V\setminus \{i\}}, 
         \end{equation}
        and select a subset of variables. Denote the index set of the selected variables by $\hat{S}_i$, which can be viewed as an estimate of $S_i$. } 
        
        \For{each pair of variables $(X_i, X_j)$, $1 \leq j < i \leq p_n$}{
     \textbf{(ii) Variable selection:}  
         conduct nonlinear/non-Gaussian regression
          \begin{equation} \label{screeneq2}
           X_j\sim  \bX_{V\setminus \{i,j\}},
           \end{equation}
           and select a subset of variables. Denote the index set of the selected variables  by  $\hat{S}_{j\setminus i}$, which can be viewed as an estimate of $S_{j\setminus i}$. 
           
     \textbf{(iii) Conditional independence test:}  
       perform nonparametric conditional independence test 
            \begin{equation} \label{test1}
             X_i \indep X_j| \bX_{\hat{S}_i \cup \hat{S}_{j\setminus i} \setminus \{j\}},
             \end{equation}
             and denote the $p$-value of the test by 
              $q_{ij}$. }
     \textbf{(iv) Multiple hypothesis test:}  perform a multiple hypothesis test for all pairs of variables based on the $p$-values $\{q_{ij}: 1 \leq j <i \leq p_n \}$, and identify the pairs for which the $p$-values are significantly smaller than others.
    \caption{Double Regression}
    \label{Alg1} 
   \end{algorithm}

  The proposed method is called double regression, as for each conditional independence test two regression tasks need to be performed in order to reduce the size of the conditioning set. Regarding this method, we have three remarks: 
  
 \begin{remark} \label{rem1} 
  When $X_i$'s are Gaussian and linearly dependent, the double regression method is reduced to the $\psi$-learning method \citep{LiangSQ2015}, 
  for which it can be shown that the $p$-value of the conditional independence test (\ref{test1}) provides an equivalent measure for the partial correlation coefficient $\rho_{ij|V\setminus \{i,j\}}$ in determining the structure of the GGM. Note that, for GGMs, the conditioning set used in (\ref{test1}) can be further reduced based on the Markov and faithfulness properties as shown in \cite{LiangSQ2015}. 
 \end{remark}
 
 \begin{remark} \label{rem2} 
 In addition to consistent variable selection procedures, (\ref{prob1}) also holds for variable sure screening procedures. 
 %For this reason, Algorithm \ref{Alg1} is described in the more general term ``variable sure screening''.  
 In practice, variable screening for high-dimensional nonlinear regression can be done using Bayesian sparse neural networks \citep{Liang2018BNN,SunSLiang2021}, which are shown to have the sure screening property for both normal and multinomial logistic regression. For other continuous variables, a nonparanormal transformation 
 \citep{LiuH2009} can be applied before the application 
 of Bayesian sparse neural networks. 
  One can also replace the Bayesian sparse neural network by a model-free sure independence screening (SIS) procedure, say, the Henze-Zirkler sure independence screening (HZ-SIS) procedure proposed by \cite{XueLiang2017} for nonlinear regression with continuous variables, and 
  the sure independence screening procedure proposed by 
  \cite{MVSIS2015} for nonlinear regression with a categorical response variable. 
  When a sure independence screening procedure is used, we recommend to select 
  $n/(c\log(n))$ variables for each regression task, where $c$ may slightly vary from 1 for different problems. 
  More interestingly, the above two types of variable screening procedures can be used in a combined manner; that is, one can first perform a SIS procedure to reduce the dimension of the conditioning set, and then perform Bayesian sparse neural networks to have the dimension of the reduced conditioning set reduced further.  
 \end{remark}
 
 \begin{remark}
 Our recommendations for the variable sure screening methods and nonparametric conditional independence tests (CIT) are summarized in Table \ref{recommended_methods}.
 
\begin{table}[htbp]
\caption{Methods recommended for variable sure screening and nonparametric conditional independence tests.}
\centering
\label{recommended_methods}
\begin{tabular}{l|l}
\toprule
 Procedure     & Recommended Methods  \\ \midrule
\begin{tabular}[c]{@{}l@{}}Variable Sure \\ Screening \end{tabular}     & \begin{tabular}[c]{@{}l@{}} Methods collected in Model-free SIS Procedures \citep{Cheng2022}, \\ 
%Distance Correlation SIS \citep{li2012feature}, \\ 
%Quantile-Adaptive SIS \citep{he2013quantile}, \\ 
Henze-Zirkler SIS \citep{XueLiang2017}, \\ 
Sparse BNN \citep{Liang2018BNN,SunSLiang2021}\end{tabular} \\ \midrule
\begin{tabular}[c]{@{}l@{}}Nonparametric Conditional \\ Independence Test (CIT)\end{tabular} & \begin{tabular}[c]{@{}l@{}}Kernel CIT \citep{zhang2012kernel}, \\ Conditional Distance Independence Test \citep{wang2015conditional},\\ Permutation-based Kernel CIT \citep{doran2014permutation}, \\ Generative CIT \citep{bellot2019conditional}\end{tabular}                             \\ \bottomrule
\end{tabular}
\end{table}
 \end{remark}

 \subsection{Consistency}
 
 This section establishes the consistency of the proposed double regression method for learning graphical models with high-dimensional nonlinear and/or non-Gaussian data. 
 Let $T_{ij}$ denote the test statistic used in the conditional independence test (\ref{test1}). 
 Let $\hat{\mathcal{E}}_n$ denote the resulting network 
 by Algorithm \ref{Alg1}. 
 To study the consistency of $\hat{\mathcal{E}}_n$, we make the following assumptions.  

 \begin{assumption} \label{ass1} (Markov and faithfulness properties)
  The generative distribution of the data is Markov and faithful to a directed acyclic graph. 
\end{assumption}

\begin{assumption} \label{ass2} (High dimensionality) The dimension $p_n$ increases in a polynomial of the sample size $n$.   
\end{assumption} 
 
 \begin{assumption} \label{ass3} (Uniform sure screening property) The variable screening procedure satisfies the uniform sure screening property, i.e., $\min_{1\leq i\leq p_n} P(S_i \subset \hat{S}_i) \to 1$ and $\min_{1\leq j<i \leq p_n} P(S_{j\setminus i} \subset \hat{S}_{j \setminus i}) \to 1$ hold as the sample size $n\to \infty$. 
 \end{assumption} 
 
 \begin{assumption} \label{ass4} (Separation)  $\min_{i,j}(\mu_{ij,1}-\mu_{ij,0})> \eta_n$, where $\eta_n=c_0 n^{-\kappa}$ for some 
 constant $c_0>0$ and $\kappa>0$, and 
 $\mu_{ij,0}$ and $\mu_{ij,1}$ denote the mean values of the test statistic $T_{ij}$ under the null (conditionally independent) and alternative (conditionally dependent) hypotheses, respectively.
 \end{assumption} 
 
\begin{assumption} \label{ass5} (Tail probability)  \[
\sup_{i,j} P(|T_{ij}-\mu_{ij}|>\frac{1}{2}\eta_n) 
=\exp\left\{-O(n^{\delta(\kappa)})\right\}, 
\]
where $\mu_{ij}=E(T_{ij})$ denotes the expectation of $T_{ij}$, and  
$\delta(\kappa)$ is positive number depending on  $\kappa$.
\end{assumption}
  
Regarding these assumptions, 
we have the following comments. Assumptions \ref{ass1} and \ref{ass2} are regular for high-dimensional graphical models. Similar assumptions have been used in the study of Gaussian graphical models, see e.g. \cite{LiangSQ2015}. Assumption \ref{ass3} ensures  asymptotic validity of the conditional independence tests on reduced conditioning sets. Under Assumption \ref{ass2}, it is easy to verify that the uniform sure screening property is satisfied by many existing sure independence screening procedures
such as those proposed in 
\cite{XueLiang2017} and \cite{MVSIS2015}. This property also holds for Bayesian sparse neural networks under Assumption \ref{ass2}. Assumptions \ref{ass4} and \ref{ass5} are on the distribution of the test statistics. Assumptions \ref{ass4} is 
like the $\beta$-min condition used in high-dimensional variable selection, see e.g. \cite{dezeure2015high} and \cite{van2011adaptive}. This condition basically requires that the gap between the mean values of the test statistics under the null and alternative hypotheses are “sufficiently large” to separate, which seems necessary for proving the consistency of Algorithm \ref{Alg1}. Assumption \ref{ass5} constrains the tail probability of the distribution. Without loss of generality, we can assume 
that $T_{ij}$ is an average-type test statistic and follows a sub-Gaussian distribution. 
Then, by the concentration inequality, it is easy to show that Assumption \ref{ass5} holds provided that 
$0<\kappa<1/2$ and $0<\delta(\kappa)<1-2 \kappa$. 

Let $A_{ij}$ denote that an error event occurs when testing
the hypotheses $H_0: e_{ij} = 0$  versus $H_1: e_{ij}=1$, where $e_{ij}=0$ and $e_{ij}=1$ denote that the variables $X_i$ and $X_j$ are conditionally independent and dependent, respectively. Thus,
\begin{equation} \label{proofeq1}
P\{\mbox{an error occurs in $\hat{\mathcal{E}}_n$}\}
= P( \cup_{i> j} A_{ij}) \leq O(p_n^2) \sup_{i > j} P(A_{ij}).
\end{equation}
Then based on the Assumption \ref{ass1} - \ref{ass5}, we have the following theorem.
  
 \begin{theorem} \label{thm:1} Suppose Assumptions \ref{ass1}-\ref{ass5} hold. For each pair of variables $(X_i,X_j)$, let $\mu_{ij,0}+\eta_n/2$ denote the critical value of the conditional independence test (\ref{test1}). Then the network estimate is consistent, i.e., 
 $P\{\mbox{an error occurs in $\hat{\mathcal{E}}_n$}\} \to 0$ as $n\to \infty$.
 \end{theorem}
 
 Theorem \ref{thm:1} establishes consistency of the double regression method. In other words, it  shows that if constructing the network with the double regression method, the probability of mistakenly adding or removing one edge converges to 0 as the sample size goes to infinity. The proof is presented in the Appendix.
  
 \section{Synthetic Examples} \label{experiments}

%\subsection{Example 1} 

 \paragraph{Example 1} We generated 100 datasets from the following nonlinear and non-Gaussian model 
 \begin{equation} \label{7nodeeq}
 \begin{split}
     X_1 & \sim Unif[-1,1], \\
     X_2 &= 6\cos(X_1)+\epsilon_2, \quad \epsilon_2 \sim Unif[-1,1], \\
     X_3 &= 5\sin(X_1)+X_2+\epsilon_3, \quad \epsilon_3 \sim N(0,1), \\
     X_4 &= 5\cos(X_3 X_6)+3 X_3+3 X_6+\epsilon_4, \quad \epsilon_4 \sim N(0,1), \\
     X_5 &= 0.05(X_2+X_6)^3+\epsilon_5, \quad \epsilon_5 \sim N(0,1), \\
     X_6 & \sim Unif[-1,1], \\
     X_7 & = 6\cos(0.2(X_3+\log(|5 X_5|+1)))+\epsilon_7, \quad \epsilon_7 \sim Unif[-1,1], \\
     X_i& \sim N(0,1), \quad i=8,9,\ldots, p_n. \\
 \end{split}
 \end{equation}
 In this example, we set the dimension $p_n=30$ and the sample size $n=400$. To goal of this example is to compare the proposed method with the existing methods, such as notears and DAG-GNN, under their ideal setting. Note that both the methods notears and DAG-GNN are developed under the low-dimensional scenario.  
 
\begin{figure}[htbp]
\centering
\includegraphics[height=2.75in,width=5.5in]{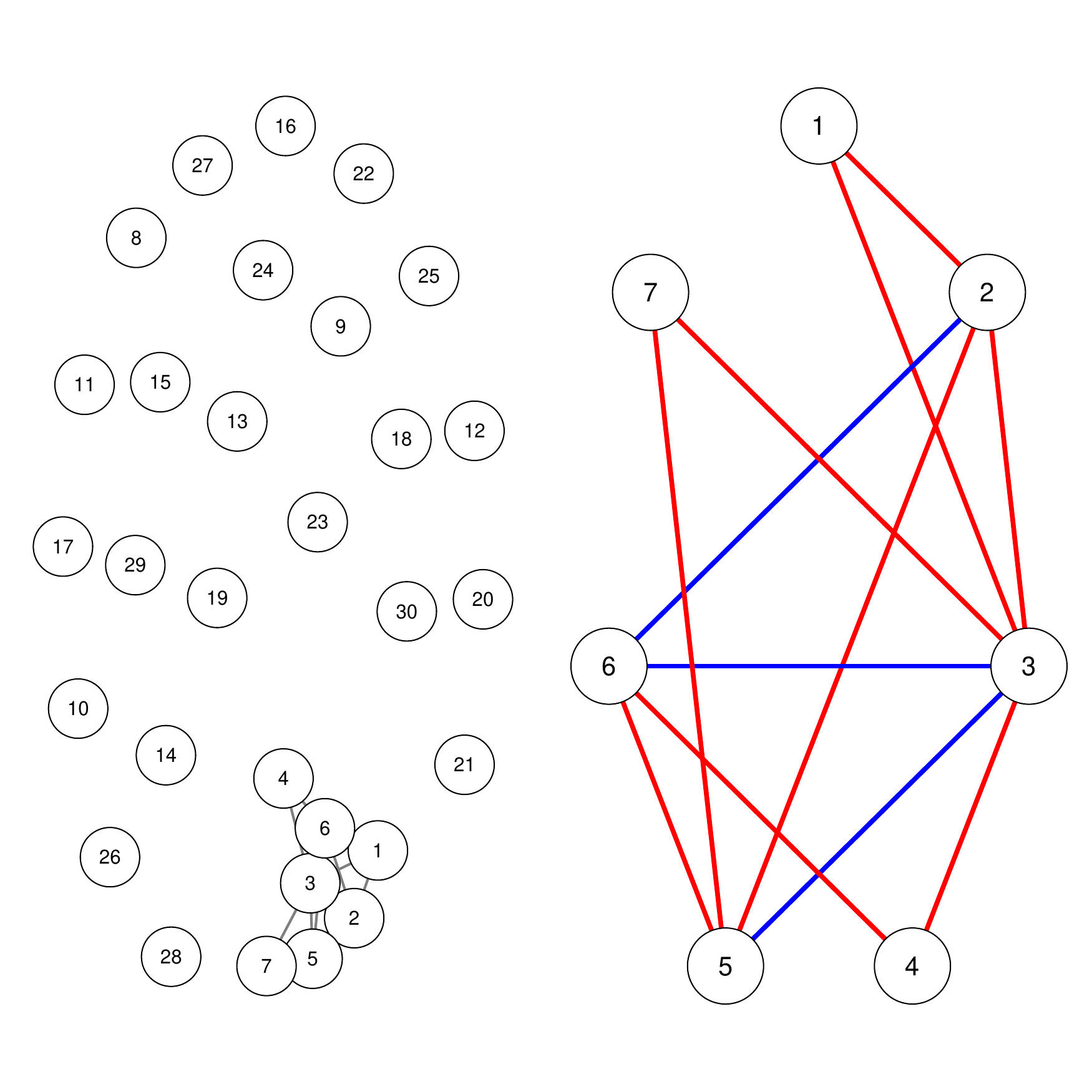}
\caption{Markov networks identified by Algorithm \ref{Alg1} for one simulated dataset of the model (\ref{7nodeeq}) at a significance level of 0.01 based on  adjusted $p$-values: (left plot) full network; 
(right plot) subnetwork of the nodes $X_1-X_7$, where the red lines indicate parents/children relationships and the blue lines indicate spouse relationships.}
\label{ex2fig}
\end{figure}

 Algorithm \ref{Alg1} was applied to this example with HZ-SIS \citep{XueLiang2017} used in variable screening 
 and the randomized conditional correlation test (RCoT) \citep{Strobl2019} used in nonparametric conditional independence test. The network edges in step (c) of Algorithm \ref{Alg1} are identified based on the adjusted $p$-values \citep{Holm1979}. 
 We use the averaged \underline{A}reas \underline{U}nder the precision-recall \underline{C}urves (AUC) as the metric to evaluate the performance of different methods.  Refer to Appendix \ref{PRdefinition} for the definitions of precision and recall. The numerical results are summarized in Figure \ref{ex2fig} and Table \ref{ex2tab}.

The left plot of Figure  \ref{ex2fig} shows the Markov network identified by Algorithm \ref{Alg1} for one simulated dataset at a significance level of $\alpha=0.01$ based on the adjusted $p$-values, where we set the neighborhood size $n_s=5$ in variable screening, i.e., $\hat{S}_i$ and $\hat{S}_{j\setminus i}$ consist of top $5$ variables which are most related to 
$X_i$ and $X_j$, respectively. For HZ-SIS, the relatedness of two variables is measured in 
the Henze-Zirkler test statistic \citep{HenzeZ1990}. 
The right plot of Figure \ref{ex2fig} shows the subnetwork of the nodes $X_1$--$X_7$, which is identical to its true Markov network of the model 
(\ref{7nodeeq}). By the standard graph theory, the undirected Markov network (also known as moral graph) corresponding to a DAG consists of all parents/children relations and spouse relations.  
The performance of the algorithm is quite robust to the choice of $\alpha$. For example, if we set $\alpha=0.05$ for this dataset, only three false links,  $X_{16}$-$X_{23}$, $X_{16}$-$X_{24}$ and $X_{19}$-$X_{27}$, are added to the full network, and the same subnetwork shown in Figure \ref{ex2fig} can still be identified.

%\begin{table} 
%\caption{Outcomes of a binary decision}
%\label{tabPR}
%\begin{center}
%\begin{tabular}{ccc} \toprule
%  &  Actual positive (P)  &  Actual negative (N) \\ \midrule
%Predicted positive  & True positive (TP) & False positive (FP) \\ 
%Predicted negative & False negative (FN) & True negative (TN) \\  \bottomrule
%\end{tabular}
%\end{center}
%\end{table}

 Table \ref{ex2tab} explores the effect of neighborhood size $n_s$ used in variable screening 
  on the performance 
  of the algorithm, where the AUC values were calculated. The precision-recall curve is often used in information retrieval for comparison of performances of different binary decision algorithms. 
  Table \ref{ex2tab} indicates that an excessively large conditioning set can deteriorate the power of the nonparametric CIT. This further implies that a direct application of a nonparametric CIT to a high-dimensional dataset for learning the Markov network is not efficient. An appropriate variable selection/screening procedure is crucial to the performance of Algorithm \ref{Alg1}. For such a low-dimensional problem, we generally recommend a variable selection method to be used in steps (i) and (ii) of Algorithm \ref{Alg1}, provided the variable selection method is not too expensive.

%  Let us define an experiment from $P$ positive instances and $N$ negative instances under some conditions. The four outcomes can be summarized in Table \ref{tabPR}. Then the precision and recall are defined by
%\[
%\mbox{precision} = \frac{TP}{TP+FP}, \quad
%\mbox{recall} = \frac{TP}{TP+FN},
%\]
%and $TP$, $FP$ and $FN$ denote true positives, false positives and false negatives, respectively.

%\begin{table}[htbp]
%\caption{Averaged areas under the precision-recall curves (AUC) produced by Algorithm \ref{Alg1}   for recovering the undirected Markov network of model (\ref{7nodeeq}), where the number in the parenthesis represents the standard deviation of the averaged AUC over 5 datasets. }
%\label{ex2tab}
%\begin{center}
%\begin{tabular}{cccc} \toprule
%Neighborhood size ($n_s$)  & 5 & 10 & 20  \\  \midrule
% Averaged AUC      & 0.9545(0.0035)  & 0.8764(0.0123) &  0.8095(0.0220) \\ \bottomrule
% \end{tabular}
% \end{center}
% \end{table}

\begin{table}[htbp]
\caption{Averaged areas under the precision-recall curves (AUC) produced by Algorithm \ref{Alg1} and baseline methods for discovering the undirected Markov network of model (\ref{7nodeeq}), where ``SD'' represents the standard deviation of the averaged AUC over 100 datasets. }
\centering
\label{ex2tab}
\begin{adjustbox}{max width=\textwidth}
\begin{tabular}{|c|c|c|c|c|c|c|c|c|c|}
\hline
\multirow{2}{*}{Method} & \multicolumn{3}{c|}{double regression}           & \multicolumn{3}{c|}{notears}                        & DAG-GNN       & GLASSO  & GSCAD \\ \cline{2-10} 
                        & 5              & 10             & 15             & linear          & MLP             & Sob             & --        & --    & --   \\ \hline
AUC                     & 0.9543 & 0.8767 & 0.8096 & 0.7935  & 0.7990  & 0.7746  & 0.7527 & 0.6346 & 0.6975 \\
 SD & 0.0008 & 0.0027 & 0.0029 & 0.0023 & 0.0018 & 0.0032 & 0.0035 & 0.0019 & 0.0022\\
% Avg. CPU time & & & & 61.1500s & 18.4050s & 3.9200s & 12.2136s \\
\hline
\end{tabular}
\end{adjustbox}
\end{table}

For comparison, we have applied the methods developed by \cite{GGMnotears_nonlinear2020} 
(denoted by ``notears'' \footnote{We use the program code available from \url{https://github.com/xunzheng/notears}.}) and \cite{Yu2019DAGGNNDS} (denoted by ``DAG-GNN'' \footnote{We use the program code available from \url{https://github.com/fishmoon1234/DAG-GNN}.}), as the baseline methods, to this example. The results are summarized in Table \ref{ex2tab}. 
The method notears include three options, ``linear'', ``MLP'' and ``Sob'', which are to approximate each regression (\ref{screeneq1}) by a linear regression, multilayer perceptron, and Sobolev basis function, respectively. The ``linear'' corresponds to the Gaussian graphical model. The comparison indicates that the double regression method can outperform the baseline methods significantly.

For a thorough comparison, we have also applied the methods developed for GGMs, such as 
Gaussian-Lasso (GLASSO) \footnote{We use the \emph{ggmncv()} function with ``lasso'' penalty in \emph{GGMncv} package available at \url{https://cran.r-project.org/web/packages/GGMncv/index.html}.} 
and Graphical-SCAD (GSCAD) \footnote{We use the \emph{ggmncv()} function with ``scad'' penalty in \emph{GGMncv} package available at \url{https://cran.r-project.org/web/packages/GGMncv/index.html}.} to this example, although the data are not Gaussian. The numerical results summarized in Table \ref{ex2tab} show that the proposed double regression method also significantly outperforms these GGM methods for this example.

% \subsection{Example 2} 
 
\paragraph{Example 2} We generated 100 datasets from the following high-dimensional nonlinear and non-Gaussian model: 
 \begin{equation} \label{AR2model}
 \begin{split}
     X_1 & \sim Unif[-1,1], \quad 
     X_2 =g(X_1)+\epsilon_2, \quad \epsilon_2\sim N(0,1)  \\ 
     X_i&=f(X_{i-2})+g(X_{i-1})+\epsilon_i, \quad i=3,4,\ldots, p_n, \\
 \end{split}
 \end{equation}
 where $\epsilon_i$ was drawn with an equal probability from $N(0,1)$ or $Unif[-0.5,0.5]$ for $i=3,4,\ldots,p_n$,
 and both $f(z)$ and $g(z)$ were randomly drawn from the set of functions $\{|z|\cos(z), \tanh(z)$, $\log(|z|+1)\}$ for $i=2,3,\ldots,p_n$. 
 In this example, we set the dimension $p_n=1000$ and the sample size $n=400$, which represents a small-$n$-large-$p$ problem. 
 
 Algorithm \ref{Alg1} was applied to this example, where variable screening was done using the HZ-SIS procedure 
 \citep{XueLiang2017}, 
  the nonparametric CIT was done using the RCoT  \citep{Strobl2019}, and the network edges in step (c) 
  were identified based on the adjusted $p$-values \citep{Holm1979} of 
  the conditional independence tests. 
 The numerical results are summarized in Table \ref{ex1tab}.  

% \begin{table}[htbp]
% \caption{Averaged areas under the precision-recall curves (AUC) produced by Algorithm \ref{Alg1}  
%  for recovering the undirected Markov network of the model (\ref{AR2model}), where the number in the parenthesis represents the standard deviation of the averaged AUC over 5 datasets. }
% \label{ex1tab}
% \begin{center}
% \begin{tabular}{ccccc} \toprule
% Neighborhood size $(n_s)$  & 5 & 10 & 25 & 50 \\  \midrule
%  Averaged AUC      & 0.8038(0.0039)  & 0.8356(0.0038) &  0.8447(0.0012) & 0.7986(0.0034) \\ \bottomrule
%  \end{tabular}
%  \end{center}
%  \end{table}

 \begin{table}[htbp]
 \caption{Averaged areas under the precision-recall (PR) curves (AUC) produced by Algorithm \ref{Alg1}  
 for recovering the undirected Markov network of the model (\ref{AR2model}), where ``SD'' represents the standard deviation of the averaged AUC over 100 datasets. }
% \vspace{-0.1in}
\label{ex1tab}
\begin{center}
\begin{adjustbox}{max width=\textwidth}
\begin{tabular}{|c|c|c|c|c|c|c|c|c|c|c|}
\hline
\multirow{2}{*}{Method} & \multicolumn{4}{c|}{double regression}                            & \multicolumn{3}{c|}{notears}      & DAG-GNN     & GLASSO  & GSCAD  \\ \cline{2-11} 
&  20  & 30  & 40  & 50  & Linear   &  MLP   & Sob    & --        & --    & --   \\ \hline
AUC  & 0.7564 & 0.7567 & 0.7556 & 0.7547 &  --  &   0.0808    &   0.3028    & 0.3296  & 0.6953 & 0.7136 \\ 
SD  & 0.0007 & 0.0008 & 0.0008 & 0.0008 & --    & 0.0024  &  0.0026  & 0.0037  & 0.0071  & 0.0073 \\
%Avg. CPU time                &        &        &        &        & $>$ 72 hrs & 27174.1599s  & 2927.3721s     & 1134.6328s\\
\hline
\end{tabular}
\end{adjustbox}
\end{center}
\end{table}

%\begin{table}[htbp]
% \caption{Averaged areas under the ROC curves (AUC) produced by Algorithm \ref{Alg1}  for recovering the undirected Markov network of the model (\ref{AR2model}), where ``SD'' represents the standard deviation of the averaged AUC over 100 datasets. }
% \vspace{-0.1in}
%\label{ex1tab2}
%\begin{center}
%\begin{adjustbox}{max width=\textwidth}
%\begin{tabular}{|c|c|c|c|c|c|c|c|c|c|c|}
%\hline
%\multirow{2}{*}{Method} & \multicolumn{4}{c|}{double regression}                            & \multicolumn{3}{c|}{notears}      & DAG-GNN     & GLASSO  & GSCAD  \\ \cline{2-11} 
%&  20  & 30  & 40  & 50  & Linear   &  MLP   & Sob    & --        & --    & --   \\ \hline
%AUC  & 0.9665 & 0.9687 & 0.9698 & 0.9720 &  --  & 0.5396      & 0.6856        & 0.6066  & 0.8798 & 0.8756 \\ 
%SD  & 0.0003 & 0.0003 & 0.0003 & 0.0002 & --    & 0.0012  & 0.0008   & 0.0029  & 0.0089  & 0.0089 \\
%Avg. CPU time                &        &        &        &        & $>$ 72 hrs & 27174.1599s  & 2927.3721s     & 1134.6328s\\
%\hline
%\end{tabular}
%\end{adjustbox}
%\end{center}
%\end{table}
 
 Table \ref{ex1tab} shows  that Algorithm \ref{Alg1} is not sensitive to neighborhood size $n_s$ under the high-dimensional setting. For this example, Algorithm \ref{Alg1} can perform reasonably well for a neighborhood size between 20 and 50.  
 Intuitively, $n_s$ cannot be too small or too large. 
 An excessively small value of $n_s$ increases the risk of missing important conditioning variables, while  an excessively large value of $n_s$ can cause the issue of ``curse of dimensionality''. They both reduces the power of the tests. 
 For such a small-$n$-large-$p$ problem, we generally recommend a variable sure screening 
 method to be used in steps (i) and (ii) of Algorithm 1.
 
% Figure \ref{ex1fig} shows the precision-recall curves produced by Algorithm \ref{Alg1} for one dataset with different neighborhood sizes. It implies that a small neighborhood size will perform well in the low recall region, while a large neighborhood size will perform well in the high recall region. %Also, the performance of the algorithm is not very sensitive to the neighborhood size. For this example, it can perform reasonably well for a neighborhood size between 10 and 25. 
 
%\begin{figure}[htbp]
%\centering
%\includegraphics[height=3.0in,width=4.75in]{figures/PRcurveARcom.pdf}
%\caption{Precision-recall curves produced by Algorithm \ref{Alg1} with different variable screening neighborhood sizes (ns).  }
%\label{ex1fig}
%\end{figure}
 
 For comparison, the baseline methods, notears, DAG-GNN, GLASSO and GSCAD, have also been applied to this example. The results are summarized in Table \ref{ex1tab}. For ``notears'', we tried all three options, linear, MLP and Sob, but the linear option caused to slow computing and did not produce any results. The comparison indicates again the superiority of the double regression method.
 Under the high-dimensional setting, the double regression method makes a drastic improvement over 
 notears and DAG-GNN, the state-of-the-art nonlinear and non-Gaussian methods.
 
% As a summary of Table \ref{ex1tab} and Figure \ref{ex1fig}, we conclude that Algorithm \ref{Alg1} performs very well for this high-dimensional nonlinear and non-Gaussian example. 

 \section{Causal Structure Discovery for High-Dimensional Regression}  \label{sect4}

\subsection{The Algorithm}

 The causal relationship for a pair or more variables refers to a {\it persistent association}
 which is expected to exist in all situations without being affected by the values of other variables. The
 causal relationship discovery has been an essential task in many disciplines.
% For high-dimensional problems, since it is usually difficult and expensive to identify causal relationships through intervention experiments, passively observed data has thus become an important source to be searched for causal relationships. The challenge of causal relationship discovery from observational data lies in the fact that statistical associations detected from observational data are not necessarily causal. 
 In statistics, the causal relationship can be determined using conditional independence tests. For a large set of variables, a pair of variables
 are considered to have no direct causal relationship if a subset of the remaining variables
 can be found such that the two variables are independent conditioned on this subset of variables. Based on conditional independence tests,
 \cite{SpirtesC2000} proposed the famous PC algorithm
 for learning the structure of causal Bayesian networks.
 Later, \cite{BuhlmannMM2010} extended
 the PC algorithm to variable selection for high-dimensional linear regression. 
 The extension is called the PC-simple algorithm which can be used to search for the causal
 structure around the response variable. 
 Note that the causal structure includes all the possible direct causes and effects of the response variable, i.e., all the parents and children in the terminology of DAGs. For certain problems, we may be able to determine in logic which are for parents and
 which are for children, although PC-simple cannot tell.
 In the same vein, \cite{LiangXJ2021} and \cite{SunLiang2021} proposed the Markov neighborhood regression (MNR) approach for high-dimensional inference and applied it to causal structure discovery around the response variable.  Like the PC-simple algorithm, MNR works based on a series of conditional independence tests. 
 
% An alternative algorithm that can be used for local causal discovery is the HITON-PC algorithm \citep{Aliferisetal2010}, which is also an extension of the PC algorithm.
 
 %The major issue with the PC-simple and HITON-PC algorithms is with their time complexity. For both algorithms, in the worst scenario, i.e., when for each of the $p$ features all conditional independence tests of order from 1 to $p-1$ are conducted, the total number of conditional tests is $O(p 2^p)$.
% Even under the sparsity constraint, the total number of conditional tests can still be of a high order polynomial of $p$. See \cite{BuhlmannMM2010} for more discussions on this issue.

 In the same logic as MNR and the PC-simple algorithm,  Algorithm \ref{Alg1} can be extended to find the causal structure around the response variable for the high-dimensional regression:
 \begin{equation} \label{regeq}
      Y=f(X_1, X_2, \ldots, X_{p_n}, \epsilon),
  \end{equation}
  where $f$ denotes a general nonlinear function, $X_1, X_2,\ldots, X_{p_n}$ are explanatory variables, and $\epsilon$ denotes random error. 
  To identify the causal variables around the response variable $Y$, 
  the extended algorithm can be described as follows. 
 
\begin{algorithm}[htbp]
\caption{Causal Structure Discovery}
\SetAlgoLined
 {\bf (i) Variable selection:}  Conduct the regression $Y \sim \bX$ to obtain a reduced feature set. 
    Denote the reduced feature set by $\hat{S} \subseteq \{1,\dots,p_n\}$ as an estimate of the set of true variables of the regression $Y \sim \bX$. 
    %and its size $|\hat{S}|$ is upper bounded by $O(\sqrt{n}/\log(n))$.

  \For{each feature $X_j \in \hat{S}$}{ 
   {\bf (ii) Variable selection:} conduct the regression  $X_j \sim \bX_{V\setminus \{j\}}$ to obtain a reduced neighborhood $\hat{\xi}_{j}\subseteq \{1,\dots,p_n\}$.
   %with the size $|\hat{\xi}_j|$ upper bounded by $O(\sqrt{n}/\log(n))$.
   
   {\bf (iii) Conditional independence test:} conduct nonparametric conditional independence test 
    \begin{equation} \label{testCS2}
     X_j \indep Y| \bX_{\hat{S} \cup \hat{\xi}_{j} \setminus \{j\}},
    \end{equation}
  and denote the $p$-value of the test by  $q_{j}$.}
  
 {\bf (iv) Multiple hypothesis test:} conduct a multiple hypothesis test to identify the causal features based on the $p$-values calculated above.
 \label{subsetAlg3}
\end{algorithm}
 
 \subsection{Drug Sensitive Gene Selection} 
 
 Disease heterogeneity is often observed in complex diseases such as cancer.
 For example, molecularly targeted cancer drugs are only
 effective for patients with tumors expressing targets \citep{GrunwaldH2003, Buzdar2009}.
 The disease heterogeneity has directly motivated the development of precision medicine,
 aiming to improve patient care by tailoring optimal therapies to an individual patient according to his/her
 molecular profile and clinical characteristics. 
 Identifying sensitive genes to different drugs
 is an important step toward the goal of precision medicine. 

 To illustrate the proposed method, we considered the CCLE dataset, which is publicly available at {\it www.broadinstitute.org/ccle}.
  The dataset consists of 8-point dose-response curves for 24 chemical compounds across over 400
 cell lines. For different chemical compounds, the numbers of cell lines are
 slightly different. For each cell line, it consists of the expression values of $p_n=18,988$ genes.  We used the area under the dose-response curve, which was termed
 as activity area in \cite{Barretinaetal2012}, to measure the sensitivity of a drug to each cell line. Compared to other measurements, such as $IC_{50}$ and $EC_{50}$, the activity area could capture the efficacy and potency of the drug simultaneously.  
 
  Algorithm \ref{subsetAlg3} was applied to the dataset collected for each drug to identify the drug-sensitive genes, where variable screening was done in two steps. First, we applied HZ-SIS \citep{XueLiang2017} to reduce the number of features for each regression to 80 ($\approx n/\log(n)$). Next, we applied the sparse Bayesian neural network (BNN) method developed by \cite{Liang2018BNN} to reduce the number of features of each regression further. 
  By \cite{Liang2018BNN}, the sparse BNN also possesses the sure screening property and thus can be used here. As a result, the sizes of $\hat{S}$ and $\hat{\xi}_j$'s can be very small, which are around 10 or less for each drug of this example. 
  With the reduced conditioning set, the accuracy of the conditional independence test (\ref{testCS2}) can be much improved. 
  The drug sensitive genes can then be identified based on the adjusted $p$-values \citep{Holm1979} of the conditional independence tests. We set the significance level of the multiple hypothesis test at 0.05.  Note that for the drug response regression, i.e., those performed in step (a) of Algorithm \ref{subsetAlg3}, the target gene of the drug will be added to the regression as an additional feature if it is not selected by HZ-SIS. 
  
  Table \ref{drugtab} summarizes the results of this example, where we validate the ``association'' between the drug and each gene by the number of PubMed articles (https://www.ncbi.nlm.nih.gov/pmc/)  which cite both the drug and the gene.  
  As shown in Table \ref{drugtab}, for many drugs, the gene selection results by Algorithm \ref{subsetAlg3} are strongly supported by relevant PubMed articles.

  For comparison, the baseline methods, notears \citep{GGMnotears_nonlinear2020} and DAG-GNN \citep{Yu2019DAGGNNDS}, and some linear model-based methods such as desparsified Lasso \citep{vandeGeer2014, Javanmard2014,ZhangZhang2014}, ridge projection, multi sample-splitting \cite{Meierhdi2016} and MNR \citep{LiangXJ2021}, were applied to this example. 
   For the linear model-based methods, since they are test-based, we selected for each drug the genes with the adjusted $p$-values less than 0.05 as significant; and if there were no genes selected at this significance level, we just reported one gene with the smallest adjusted $p$-value. For each drug, desparsified Lasso is simply inapplicable due to the ultra-high dimensionality of the dataset; the package {\it hdi} \citep{Meierhdi2016} aborted due to the excess of memory limit. Due to the same issue, {\it hdi} also aborted for some drugs when performing ridge regression. Multi sample-splitting and MNR work reasonably well for this example, with results partially overlapped with those by double regression.  
  
  For notears (with the MLP option) and DAG-GNN, we applied them to the reduced dataset by HZ-SIS \citep{XueLiang2017} as those used by double regression. Since the two methods are not test-based, it is possible that no gene is selected for some drugs.

\begin{scriptsize}
\begin{center}
\begin{longtable}{ccccccc}
\caption{Comparison of drug sensitive genes selected by double regression, ridge projection,
  multi sample-splitting (multi-split) and MNR for 24 anti-cancer drugs, where the superscripts $^*$, $\dag$ 
  and $\ddag$ denote that the numbers of relevant PubMed articles are in the ranges ``1--9'', ``10--99'' and ``100 and above'', respectively. For double regression, the number in the parentheses represents the adjusted $p$-value of the selected gene. The gene selection results of ridge projection, multi-split and MNR were taken from \cite{LiangXJ2021}.} 
\label{drugtab} \\ \toprule 
Drug & Double Regression & Ridge & Multi-Split  & MNR & notears & DAG-GNN\\ \hline
\endfirsthead

\multicolumn{7}{c}%
{{\bfseries \tablename\ \thetable{} -- continued from previous page}} \\
\hline Drug & Double Regression & Ridge  & Multi-Split  & MNR & notears & DAG-GNN\\ \hline
\endhead

\hline \multicolumn{7}{|r|}{{Continued on next page}} \\ \hline
\endfoot
\endlastfoot
%%%%%
17-AAG&\makecell{NQO1$^{\ddag}$ (2.16e-6)\\ ZFP30$^*$ (8.80e-4)}
&\makecell{--} &\makecell{NQO1$^\ddag$}&\makecell{NQO1$^\ddag$} & \makecell{NQO1$^\ddag$} & \makecell{NQO1$^\ddag$\\ATP6V0E1}\\\hline 
%%%%
AEW541&\makecell{IGF1R$^\ddag$(8.02e-5)\\ SGPP1$^*$(6.58e-4)\\ SP1$^\dag$(1.76e-2)}  &\makecell{F3$^\dag$}&\makecell{SP1$^\dag$}&\makecell{TMEM229B}& \makecell{IGF1R$^\ddag$} & \makecell{PUM2}\\\hline 
%%%
AZD0530&\makecell{DDAH2(2.16e-2)\\ FGFBP1(2.23e-2)} &\makecell{PPY2}&\makecell{SYN3}&\makecell{DDAH2} & \makecell{--} & \makecell{--} \\\hline 
%%%%
AZD6244&\makecell{CSF1$^\dag$(6.39e-3)\\  SPRY2$^\dag$(1.64e-2)\\CAPNS2(1.64e-2)}&\makecell{OSBPL3}&\makecell{SPRY2$^\dag$\\LYZ$^*$\\RNF125$^*$}&\makecell{LYZ$^*$\\SPRY2$^\dag$} & \makecell{SPRY2$^\dag$} & \makecell{SPRY2$^\dag$\\LYZ$^*$\\RNF125$^*$}\\\hline 
%%%%
Erlotinib&\makecell{MGC4294(2.18e-3)\\ EGFR$^\ddag$(2.35e-3) }&\makecell{LRRN1$^*$}&\makecell{PCDHGC3$^*$}&\makecell{ENPP1$^\dag$} & \makecell{--} & \makecell{--}\\\hline 
%%%%
Irinotecan&\makecell{SLFN11$^\ddag$(4.50e-11)\\CD63$^\ddag$(1.19e-6)} &\makecell{SLFN11$^\ddag$}&\makecell{ARHGAP19\\SLFN11$^\ddag$}&\makecell{ARHGAP19\\SLFN11$^\ddag$} & \makecell{ARHGAP19\\SLFN11$^\ddag$} & \makecell{SLFN11$^\ddag$\\CD63$^\ddag$\\(+5 insensitive genes)}\\\hline 
%%%%
L-685458&\makecell{GSS(2.81e-2)\\CTSL1(3.78e-2)} &\makecell{--}&\makecell{MSL2}&\makecell{FAM129B} & \makecell{RGS18} & \makecell{RGS7BP}\\\hline 
%%%
Lapatinib&\makecell{GRB7$^\ddag$(1.56e-3)\\SYTL1(8.40e-3)} &\makecell{WDFY4}&\makecell{ERBB2$^\ddag$}&\makecell{SYTL1} & \makecell{--} & \makecell{--}\\\hline 
%%%
LBW242&\makecell{RIPK1$^\dag$(3.78e-3)\\ TMEM177(3.14e-2)} &\makecell{RXFP3}&\makecell{LOC100009676}&\makecell{RIPK1$^\dag$} & \makecell{RIPK1$^\dag$} & \makecell{--}\\ \hline 
%%%
Nilotinib &\makecell{RHOC$^\dag$(1.56e-2)\\ ARID1B$^*$(1.56e-2)}&\makecell{--}&\makecell{RAB37$^*$}&\makecell{RHOC$^\dag$} & \makecell{ABL1$^\ddag$\\CNOT7\\RABL5} & \makecell{--}\\\hline 
%%%
Nutlin-3&\makecell{CCDC30(4.63e-3)\\ LOC285548(4.30e-2)\\ZMAT3$^\dag$(4.30e-2)}&\makecell{TTC7B}&\makecell{LOC100009676}&\makecell{DNAJB14} & \makecell{--}& \makecell{--} \\\hline
%%%
Paclitaxel&\makecell{BCL2L1$^\ddag$(4.39e-4)\\ SSRP1$^\dag$(1.27e-2)\\ SLC35F5$^*$(2.89e-2)\\ ZNRD1$^\dag$(3.35e-2)}&\makecell{ABCB1$^\ddag$}&\makecell{ABCB1$^\ddag$}&\makecell{BCL2L1$^\ddag$} & \makecell{ABCB1$^\ddag$}& \makecell{BCL2L1$^\ddag$\\ SSRP1$^\dag$\\(+8 insensitive genes)}\\\hline 
%%%
Panobinostat&\makecell{EIF4EBP2$^*$(2.94e-3)\\ LARP6(2.94e-3)\\ AXL$^\ddag$(2.41e-2)\\ TGFB2$^\dag$(3.76e-2)} &\makecell{C17orf105}&\makecell{PUM2}&\makecell{TGFB2$^\dag$} & \makecell{--} & \makecell{EIF4EBP2$^*$\\ AXL$^\ddag$,\\ MYB$^*$ \\PARP1$^*$\\(+4 insensitive genes)} \\\hline 
%%%%
PD-0325901&\makecell{SPRY2$^\dag$(3.95e-3)\\KLF3$^*$(1.60e-2)}&\makecell{ZNF646}&\makecell{LYZ*\\RNF125}&\makecell{DBN1} & \makecell{SPRY2$^\dag$} & \makecell{ETV4\\CRIM1\\CYR61}\\\hline
%%%%
PD-0332991&\makecell{COX18(6.97e-4)\\NTN4$^*$(4.03e-2)}&\makecell{GRM6}&\makecell{LOC100506972}&\makecell{PUM2} & \makecell{--}&\makecell{LOC100506779}\\\hline 
%%%%
PF2341066&\makecell{HGF$^\ddag$(7.06e-3)\\ ENAH$^*$(7.27e-3)\\ LRMP$^*$(1.53e-2)} &\makecell{WDFY4$^*$}&\makecell{SPN$^\dag$}&\makecell{HGF$\ddag$ \\ENAH*\\GHRLOS2} & \makecell{HGF$^\ddag$\\CBFA2T3}& \makecell{--}\\\hline 
%%%%
PHA-665752&\makecell{INHBB$^*$(1.00e-2)} &\makecell{--}&\makecell{LAIR1}&\makecell{INHBB$^*$} & \makecell{--}& \makecell{--}\\\hline 
%%%%
PLX4720&\makecell{PLEKHH3(8.08e-5)\\ MEX3C(1.78e-4)} &\makecell{ADAMTS13}&\makecell{SPRYD5}&\makecell{PLEKHH3} & \makecell{PLEKHH3\\PSORS1C1}& \makecell{GYPC}\\\hline 
%%%
RAF265&\makecell{SYT7(5.63e-3)\\PIK3CD$^*$(2.12e-2)}&\makecell{LOC100507235}&\makecell{SIGLEC9}&\makecell{SEPT11} & \makecell{--} & \makecell{--}\\\hline 
%%%
Sorafenib&\makecell{RPL22$^\dag$(4.34e-2)}&\makecell{--}&\makecell{SBNO1}&\makecell{RPL22$^\dag$ \\LAIR1$^*$} & \makecell{--} & \makecell{--}\\\hline 
%%%
TAE684&\makecell{SP1$^\dag$(1.42e-3)\\ MBNL3(1.99e-2)} &\makecell{--}&\makecell{ARID3A}&\makecell{ARID3A} & \makecell{ALK$^\dag$\\LAIR1} & \makecell{ALK$^\dag$\\TNFRSF12A\\MYOF}\\\hline
%%%
TKI258&\makecell{MYO5B(4/49e-4)\\ FECH(4.74e-2)}  &\makecell{--}&\makecell{SPN$^*$}&\makecell{KHDRBS1} & \makecell{--} & \makecell{C4orf46\\ORC1}\\\hline 
%%%%
Topotecan&\makecell{SLFN11$^\ddag$(3.73e-11)\\ CD63$^\dag$(9.46e-5)\\ HSPB8$^*$(3.59e-2)}&\makecell{--}&\makecell{SLFN11$^\ddag$}&\makecell{SLFN11$^\ddag$} & \makecell{SLFN11$^\ddag$} & \makecell{SLFN11$^\ddag$\\ HSPB8$^*$\\ DSP}\\\hline 
%%%%
ZD-6474&\makecell{APOO(9.87e-4)\\KLF2$^*$(4.59e-3)}&\makecell{MID1IP1}&\makecell{NOD1}&\makecell{PXK} & \makecell{--} & \makecell{--}\\ \bottomrule
%%%%%%%%
\end{longtable}
\end{center}
\end{scriptsize}

  Table \ref{drugtab} summarizes the results of all the above methods. It shows that the double regression and its competitors can produce similar or overlapped results for some drugs, while, in general, the gene selection results by double regression are more supported by the existing literature. For example, for the drugs Topotecan and Irinotecan, double regression, notears, DAG-GNN, multi-split and MNR all selected the gene SLFN11 as a drug sensitive gene. In the literature, \cite{Barretinaetal2012} and \cite{Zoppolietal2012} reported that SLFN11 is predictive of treatment response for Topotecan and Irinotecan.
  For the drug 17-AAG, five methods selected NQO1 as a drug sensitive gene. 
  In the literature, \cite{HadleyH2014} and \cite{Barretinaetal2012} reported NQO1 as the top predictive biomarker for 17-AAG.
  For the drug Paclitaxel, double regression, MNR and DAG-GNN selected BCL2L1 as a drug sensitive gene. In the literature, many publications, such as \cite{LeeHLYK2016} and
   \cite{Domanetal2016}, reported that the gene BCL2L1 is predictive of treatment response for Paclitaxel. For the drug PF2341066, \cite{LawrenceSalgia2010} reported that HGF, which
 was selected by double regression, MNR and notears as the drug sensitive gene, is potentially responsible for the effect of PF2341066.
 For drug LBW242, RIPK1 was selected by double regression, MNR and notears. In \cite{Gaither2007} and \cite{Moriwaki2015}, it was stated that RIPK1 is one of the presumed target of LBW242, which is involved in increasing death of cells. 
 
 For many other drugs, double regression produced more accurate gene selection results than its competitors. For example, for the drug Erlotinib whose target gene is EGFR, double regression selected its target gene as the drug-sensitive gene while its competitors did not. 
  For the drug Lapatinib, double regression selected 
  GRB7, while its competitors did not. \cite{Nencioni2010} reported that the removal of GRB7 by RNA-interference reduces breast cancer cell viability and increases the activity of Lapatinib, and they further concluded that GRB7 upregulation is a potentially adverse consequence of HER2 signaling inhibition, and preventing GRB7 accumulation and/or its interaction with receptor tyrosine kinases may increase the benefit of HER2-targeting drugs. 
 
 In summary, this example shows that the double regression method can lead to more accurate discoveries for gene regulatory relationships than the existing methods. 
 
 % Consistent literature can also be found for the gene selection results for some
 %other drugs, such as Lapatinib \citep{Nencioni2010},
 %AZD6244 \citep{Holt2012} and TAE684 \citep{Liuetal2010}.
 %\cite{Nencioni2010} reported that the removal of GRB7
 % by RNA-interference reduces breast cancer cell viability and increases the activity of Lapatinib, and
 %they concluded that GRB7 upregulation is a potentially adverse consequence of HER2 signaling inhibition,
 %and preventing GRB7 accumulation and/or its interaction with receptor tyrosine kinases
 %may increase the benefit of HER2-targeting drugs.
 % \cite{Holt2012} reported that DUSP6 is one of the key genes under MEK function control,
 %while MEK is the target gene of AZD6244.
 %\cite{Liuetal2010} reported that ARID3A plays important roles in embryonic patterning,
 %cell lineage gene regulation, and cell cycle control, chromatin remodeling
 %and transcriptional regulations, and has
 %potential values in future drug discovery and design.

 \section{Computational Complexity Analysis} \label{computationalcomplexanalysis}
 
 The computation of Algorithm \ref{Alg1} can be greatly simplified. In practice, we can construct a DAG based on the result of step (i)-(iii) by treating $\hat{S}_i$ as the set of parent nodes of $X_i$, and then establish a moral graph for the DAG. Denote the moral graph by $\hat{\mathcal{G}}_n$ and the true one by $\mathcal{E}_n$. It is easy to show by the variable sure screening property that $P(\mathcal{E}_n \subset \hat{\mathcal{G}}_n) \to 1$ as $n\to \infty$.  Therefore, the conditional independence test (\ref{test1}) can only be performed on the links of $\hat{\mathcal{G}}_n$ instead of all possible links as prescribed in Algorithm  \ref{Alg1}. Let $l_{n,p_n}$ denote the number of links contained in the true 
 sparse graph. Then the computational complexity of Algorithm \ref{Alg1} is $O((p_n +l_{n,p_n})T(n,p_n))$, where $T(p_n,n)$ denotes the computational complexity of a single regression task, and the factors $p_n$ and $l_{n,p_n}$ are for the computational complexities of steps (i) and  (ii), respectively.  
 It is reasonable to assume that the true graph has a sparsity of $l_{n,p_n}=O(p_n)$. 
 Therefore, Algorithm \ref{Alg1} has a computational complexity of $O(p_n T(n,p_n))$.
 
 We note that for both baseline methods \cite{GGMnotears_nonlinear2020} and \cite{Yu2019DAGGNNDS}, they also need to perform $p_n$ nonlinear regression essentially as implied by their loss functions. That is, all three methods have about the same 
 computational complexity. However, this comparison is somewhat unfair to 
 the proposed algorithm, as both baseline methods require the learned nonlinear regression for each node to be consistent such that the subsequent inference for 
 conditional independence relationships can be made. Under their current setting, consistency 
 might not hold for the small-$n$-large-$p$ problems, while the proposed algorithm is particularly designed for small-$n$-large-$p$ problems.  This is consistent with our numerical results, see Tables
 \ref{ex2tab} and \ref{ex1tab} for the low- and high-dimensional cases, respectively. 
 
 If the computational complexity $T(n,p_n)$ for a single regression task is considered high for some problems, the double regression for each conditional tests can be avoided. In this case, steps (ii) and (iii) of Algorithm \ref{Alg1} can be replaced by the following steps: 
 
 \vspace{-0.1in} 
 
 \begin{itemize}
     \item[(ii$^{\prime}$)] {\it For each node $X_i$, 
find the spouse nodes,  
i.e., finding the set  $A_i=\{j: \exists \  k \in \hat{S}_i \cap \hat{S}_j\}$ for $i=1,\ldots, p_n$,
  where $X_j$ is a node not connected but sharing a common neighbor with $X_i$. Let $\tilde{M}_i=\hat{S}_i \cup A_i$.} 
 \item[(iii$^{\prime}$)] {\it For each pair of variables $(X_i,X_j)$, perform nonparametric conditional independence test 
            \begin{equation} \label{test2}
             X_i \indep X_j| \bX_{\tilde{M}_{ij} \setminus \{i,j\}},
             \end{equation}
    where $\tilde{M}_{ij}=\tilde{M}_i$ if 
    $|\tilde{M}_i\setminus \{i,j\}| \leq |\tilde{M}_j \setminus \{i,j\}|$ and $\tilde{M}_{ij}=\tilde{M}_j$ otherwise. Denote the $p$-value of the test by 
     $q_{ij}$. }
 \end{itemize}
 
 The modified algorithm can be justified based on the theory of DAG. For each node $X_i$, step (i) of Algorithm \ref{Alg1} is to find the set of parents and children nodes,
  step (ii$^{\prime}$) is to find the set of spouse nodes, 
 and then $\tilde{M}_i$ forms a super Markov blanket 
 of $X_i$. Recall that the Markov blanket of a node
 in a DAG is the union of its parents, children, and spouse nodes.
 %Further, the consistency of Algorithm \ref{Alg2} is obvious by Theorem \ref{them1} under Assumptions \ref{ass1}-\ref{ass5}. Here we have assumed that the test (\ref{test2}) also satisfies Assumptions \ref{ass4}-\ref{ass5}. 
 Then, in a similar way to Theorem 1 of \cite{XuJiaLiang2019}, we will be able to show that the tests (\ref{test1}) and (\ref{test2}) are equivalent in determining the Markov network in the sense  
\[
  \phi_{ij}=1 \Longleftrightarrow \tilde{\phi}_{ij}=1,
 \]
 where $\phi_{ij}\in\{0,1\}$ and $\tilde{\phi}_{ij}\in\{0,1\}$ denote the 
 outputs of the conditional independence tests (\ref{test1}) and (\ref{test2}), respectively. 
 Again, as argued earlier, the conditional independence test (\ref{test2}) can only be performed 
 for the links of $\hat{\mathcal{G}}_n$ only, which can have the number of tests reduced significantly. 
 
\section{Discussion}\label{disucssion}

 In this paper, we have proposed an effective method for learning graphical models for high-dimensional nonlinear and/or non-Gaussian data. The proposed method works by performing a series of nonparametric conditional independence tests, for each of which the conditioning set is reduced by a double regression procedure. Since variable screening is usually much cheaper in computation than variable selection for high-dimensional regression, variable screening is recommended 
 in general. Moreover, when a sure independence variable screening procedure is employed in Algorithm \ref{Alg1}, the procedure only needs to be applied to each variable once as the results of (\ref{screeneq2}) can be directly derived from those of (\ref{screeneq1}) in this case. 
 
 Finally, we note that despite its simplicity, the conditional distribution property (\ref{prob1}) provides an alternative justification for Markov neighborhood regression \citep{LiangXJ2021, SunLiang2021} and 
 has many potential applications in high-dimensional statistical statistics. 

% \newpage

\appendix
\section{Appendix}
%You may include other additional sections here.

\subsection{Proof of Theorem 1}
 
 \begin{proof} 
%  Let $A_{ij}$ denote that an error event occurs when testing
% the hypotheses $H_0: e_{ij} = 0$  versus $H_1: e_{ij}=1$, where $e_{ij}=0$ and $e_{ij}=1$ denote that the variables $X_i$ and $X_j$ are conditionally independent and dependent, respectively. Thus,
% \begin{equation} \label{proofeq1}
% P\{\mbox{an error occurs in $\hat{\mathcal{E}}_n$}\}
% = P( \cup_{i> j} A_{ij}) \leq O(p_n^2) \sup_{i > j} P(A_{ij}).
% \end{equation}
Let $A_{ij}^I$ and $A_{ij}^{II}$ denote the false positive and false negative errors, respectively. Then,
\begin{equation} \label{proofeq2}
A_{ij} = A_{ij}^I \cup A_{ij}^{II},
\end{equation}
where
\[
\begin{cases}
  \mbox{False positive error } A_{ij}^I, &  T_{ij}>\mu_{ij,0}+\eta_n/2 \  \mbox{and} \ e_{ij}=0, \\
  \mbox{False negative error $A_{ij}^{II}$}, & T_{ij}\leq \mu_{ij,0}+\eta_n/2 \ \mbox{and} \ e_{ij}=1, \\
\end{cases}
\]
Then there exists a constant $C>0$ such that 
\begin{equation} \label{proofeq3}
\sup_{i,j} P(A_{ij}^I)= \sup_{i,j} P(T_{ij} > \mu_{ij,0}+\eta_n/2) \leq \sup_{i,j} P(|T_{ij}-\mu_{ij,0}|>\eta_n/2) \leq  e^{-c_1 n^{\delta(c_0,\kappa)}},
\end{equation}
for some positive constant $c_1$. Further, under the alternative hypothesis $H_1: e_{ij}=1$, 
\begin{equation} \label{proofeq4}
\sup_{i,j} P(A_{ij}^{II})= \sup_{i,j} P(T_{ij} \leq  \mu_{ij,0}+\eta_n/2) \leq \sup_{i,j} P(|T_{ij}-\mu_{ij,1}|>\eta_n/2) \leq  e^{-c_1 n^{\delta(c_0,\kappa)}}.
\end{equation}
Summarizing (\ref{proofeq1})-(\ref{proofeq4}), we 
have $P\{\mbox{an error occurs in $\hat{\mathcal{E}}_n$}\} \to 0$ as $n\to \infty$ under  Assumption \ref{ass2}. This concludes the proof.
\end{proof}

\subsection{Definition of Precision and Recall}  \label{PRdefinition}

Let us define an experiment from $P$
positive instances and $N$ negative instances under some conditions. The four outcomes can be summarized in 
Table \ref{tabPR}. 

\begin{table}[htbp] 
\caption{Outcomes of a binary decision}
\label{tabPR}
\begin{center}
\begin{tabular}{ccc} \toprule
  &  Actual positive (P)  &  Actual negative (N) \\ \midrule
Predicted positive  & True positive (TP) & False positive (FP) \\ 
Predicted negative & False negative (FN) & True negative (TN) \\  \bottomrule
\end{tabular}
\end{center}
\end{table}

Then the precision and recall are defined by
\[
\mbox{precision} = \frac{TP}{TP+FP}, \quad
\mbox{recall} = \frac{TP}{TP+FN},
\]
and $TP$, $FP$ and $FN$ denote true positives, false positives and false negatives, respectively.

%\bibliographystyle{asa}
%\bibliography{ref}

\end{document}